%% file: sigir2020.tex
\documentclass[sigconf]{acmart}

\usepackage{booktabs} %
\usepackage{soul}

\setcopyright{rightsretained}

\usepackage{algorithm}
\usepackage{multirow}
\usepackage{tabularx}
\usepackage{footmisc}
\newcolumntype{Y}{>{\centering\arraybackslash}X}
\usepackage{relsize}    %
\newcommand{\sys}{ColBERT}
\newcommand{\BertBase}{BERT$_{\textnormal{base}}$}
\newcommand{\BertLarge}{BERT$_{\textnormal{large}}$}
\usepackage{amssymb}

\newcommand{\etal}{\textit{et al.}}
\newcommand{\secref}[1]{\S\ref{#1}}

\usepackage{subfig}
\usepackage{nicefrac}
\usepackage{pifont}

\usepackage{graphicx}   %

\copyrightyear{2020}
\acmYear{2020}
\setcopyright{acmlicensed}
\acmConference[SIGIR '20]{Proceedings of the 43rd International ACM SIGIR Conference on Research and Development in Information Retrieval}{July 25--30, 2020}{Virtual Event, China}
\acmPrice{15.00}
\acmDOI{10.1145/3397271.3401075}
\acmISBN{978-1-4503-8016-4/20/07}

\begin{document}
\fancyhead{} %
\title{\sys{}: Efficient and Effective Passage Search via Contextualized Late Interaction over BERT}

\author{Omar Khattab}
\affiliation{
  \institution{Stanford University}
}
\email{okhattab@stanford.edu}

\author{Matei Zaharia}
\affiliation{
  \institution{Stanford University}
}
\email{matei@cs.stanford.edu}

\renewcommand{\shortauthors}{Authors}

\begin{abstract}
Recent progress in Natural Language Understanding (NLU) is driving fast-paced advances in Information Retrieval (IR), largely owed to fine-tuning deep language models (LMs) for document ranking. While remarkably effective, the ranking models based on these LMs increase computational cost by orders of magnitude over prior approaches, particularly as they must feed each query--document pair through a massive neural network to compute a single relevance score. To tackle this, we present \sys{}, a novel ranking model that adapts deep LMs (in particular, BERT) for efficient retrieval. \sys{} introduces a \textit{late interaction} architecture that independently encodes the query and the document using BERT and then employs a cheap yet powerful interaction step that models their fine-grained similarity. By delaying and yet retaining this fine-granular interaction, \sys{} can leverage the expressiveness of deep LMs while simultaneously gaining the ability to pre-compute document representations offline, considerably speeding up query processing. Beyond reducing the cost of re-ranking the documents retrieved by a traditional model, \sys{}'s \textit{pruning-friendly} interaction mechanism enables leveraging vector-similarity indexes for end-to-end retrieval directly from a large document collection. We extensively evaluate \sys{} using two recent passage search datasets. Results show that \sys{}'s effectiveness is competitive with existing BERT-based models (and outperforms every non-BERT baseline), while executing two orders-of-magnitude faster and requiring four orders-of-magnitude fewer FLOPs per query.

\end{abstract}

\maketitle

\input{1-intro.tex}

\input{2-related}

\input{3-ColBERT}

\input{4-experiments}

\input{5-conclusion}

\bibliographystyle{ACM-Reference-Format}
\bibliography{bibliography}

\end{document}

%% file: 1-intro.tex
\begin{figure}[t]
\centering
\includegraphics[width=\columnwidth]{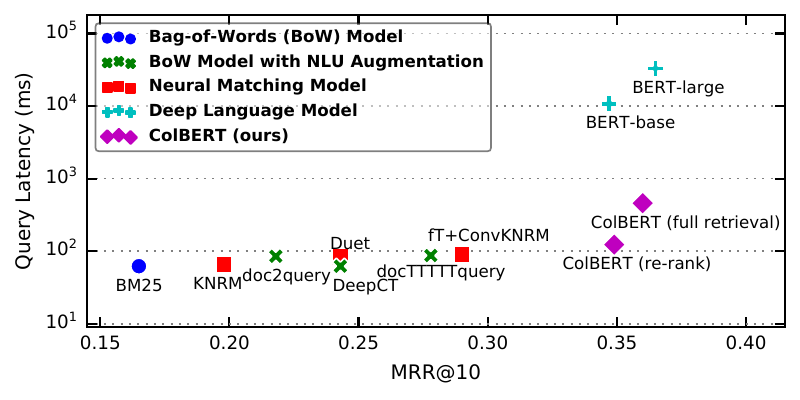}
\vspace{-7mm}
\caption{Effectiveness (MRR@10) versus Mean Query Latency (log-scale) for a number of representative ranking models on MS MARCO Ranking~\cite{nguyen2016ms}. The figure also shows \sys{}. Neural re-rankers run on top of the official BM25 top-1000 results and use a Tesla V100 GPU. Methodology and detailed results are in \secref{sec:evaluation}.}

\label{fig:tradeoff}
\vspace{-5mm}
\end{figure}

\begin{figure*}[t]
\centering
\includegraphics[width=\textwidth]{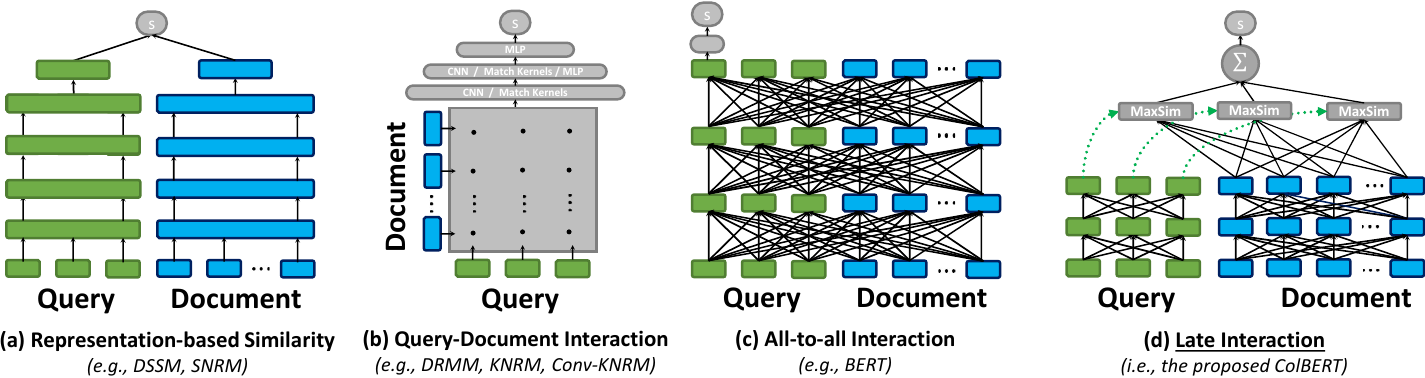}
\vspace{-5mm}
\caption{Schematic diagrams illustrating query--document matching paradigms in neural IR. The figure contrasts existing approaches (sub-figures (a), (b), and (c)) with the proposed late interaction paradigm (sub-figure (d)).}
\label{fig:X}
\vspace{-3mm}
\end{figure*}

\section{Introduction}
\label{sec:introduction}

Over the past few years, the Information Retrieval (IR) community has witnessed the introduction of a host of neural ranking models, including DRMM~\cite{guo2016deep}, KNRM~\cite{xiong2017end, dai2018convolutional}, and Duet~\cite{mitra2017learning, mitra2019updated}. In contrast to prior learning-to-rank methods that rely on hand-crafted features, these models employ embedding-based representations of queries and documents and directly model \textit{local interactions} (i.e., fine-granular relationships) between their contents. Among them, a recent approach has emerged that \textit{fine-tunes} deep pre-trained language models (LMs) like ELMo~\cite{peters2018deep} and BERT~\cite{devlin2018bert} for estimating relevance. By computing deeply-contextualized semantic representations of query--document pairs, these LMs help bridge the pervasive vocabulary mismatch \cite{zhao2012modeling, mitra2018introduction} between documents and queries~\cite{qiao2019understanding}. Indeed, in the span of just a few months, a number of ranking models based on BERT have achieved state-of-the-art results on various retrieval benchmarks \cite{nogueira2019passage, macavaney2019cedr, dai2019deeper, yilmaz2019cross} and have been proprietarily adapted for deployment by Google\footnote{https://blog.google/products/search/search-language-understanding-bert/} and Bing\footnote{https://azure.microsoft.com/en-us/blog/bing-delivers-its-largest-improvement-in-search-experience-using-azure-gpus/}.

However, the remarkable gains delivered by these LMs come at a steep increase in computational cost. Hofst\"{a}tter~\etal{}~\cite{hofstatter2019let} and MacAvaney~\etal{}~\cite{macavaney2019cedr} observe that BERT-based models in the literature are 100-1000$\times$ more computationally expensive than prior models---some of which are arguably \textit{not} inexpensive to begin with \cite{ji2019efficient}. This quality--cost tradeoff is summarized by Figure~\ref{fig:tradeoff}, which compares two BERT-based rankers \cite{nogueira2019passage, nogueira2019multi} against a representative set of ranking models. The figure uses MS MARCO Ranking~\cite{nguyen2016ms}, a recent collection of 9M passages and 1M queries from Bing's logs. It reports retrieval effectiveness (MRR@10) on the official validation set as well as average query latency (log-scale) using a high-end server that dedicates one Tesla V100 GPU per query for neural re-rankers. Following the \textit{re-ranking} setup of MS MARCO, \sys{} (re-rank), the Neural Matching Models, and the Deep LMs re-rank the MS MARCO's official top-1000 documents per query. Other methods, including \sys{} (full retrieval), directly retrieve the top-1000 results from the entire collection.

As the figure shows, BERT considerably improves search precision, raising MRR@10 by almost 7\% against the best previous methods; simultaneously, it increases latency by up to tens of thousands of milliseconds even with a high-end GPU. This poses a challenging tradeoff since raising query response times by as little as 100ms is known to impact user experience and even measurably diminish revenue \cite{bing_kdd13}. To tackle this problem, recent work has started exploring using Natural Language Understanding (NLU) techniques to augment traditional retrieval models like BM25~\cite{robertson1995okapi}. For example, Nogueira~\etal{}~\cite{nogueira2019document, nogueiradoc2query} expand documents with NLU-generated queries before indexing with BM25 scores and Dai \& Callan~\cite{dai2019context} replace BM25's term frequency with NLU-estimated term importance. Despite successfully reducing latency, these approaches generally reduce precision substantially relative to BERT.

To reconcile efficiency and contextualization in IR, we propose \textbf{\sys{}}, a ranking model based on \textbf{co}ntextualized \textbf{l}ate interaction over \textbf{BERT}. As the name suggests, \sys{} proposes a novel \textit{late interaction} paradigm for estimating relevance between a query $q$ and a document $d$. Under late interaction, $q$ and $d$ are separately encoded into two sets of contextual embeddings, and relevance is evaluated using cheap and \textit{pruning-friendly} computations between both sets---that is, fast computations that enable ranking without exhaustively evaluating every possible candidate.

Figure~\ref{fig:X} contrasts our proposed late interaction approach with existing neural matching paradigms. On the left, Figure~\ref{fig:X}~(a) illustrates \textit{representation-focused} rankers, which independently compute an embedding for $q$ and another for $d$ and estimate relevance as a single similarity score between two vectors \cite{huang2013learning,zamani2018neural}. Moving to the right, Figure~\ref{fig:X}~(b) visualizes typical \textit{interaction-focused} rankers. Instead of summarizing $q$ and $d$ into individual embeddings, these rankers model word- and phrase-level relationships across $q$ and $d$ and match them using a deep neural network (e.g., with CNNs/MLPs \cite{mitra2017learning} or kernels \cite{xiong2017end}). In the simplest case, they feed the neural network an \textit{interaction matrix} that reflects the similiarity between every pair of words across $q$ and $d$. Further right, Figure~\ref{fig:X}~(c) illustrates a more powerful interaction-based paradigm, which models the interactions between words \textit{within} as well as \textit{across} $q$ and $d$ at the same time, as in BERT's transformer architecture \cite{nogueira2019passage}.

These increasingly expressive architectures are in tension. While interaction-based models (i.e., Figure~\ref{fig:X}~(b) and (c)) tend to be superior for IR tasks \cite{guo2019deep, mitra2018introduction}, a representation-focused model---by isolating the computations among $q$ and $d$---makes it possible to pre-compute document representations offline \cite{zamani2018neural}, greatly reducing the computational load per query. In this work, we observe that the fine-grained matching of interaction-based models and the pre-computation of document representations of representation-based models can be combined by retaining yet judiciously \textit{delaying} the query--document interaction. Figure~\ref{fig:X}~(d) illustrates an architecture that precisely does so. As illustrated, every query embedding interacts with all document embeddings via a MaxSim operator, which computes maximum similarity (e.g., cosine similarity), and the scalar outputs of these operators are summed across query terms. This paradigm allows \sys{} to exploit deep LM-based representations while shifting the cost of encoding documents offline and amortizing the cost of encoding the query once across all ranked documents. Additionally, it enables \sys{} to leverage vector-similarity search indexes (e.g., \cite{JDH17,abuzaid2019index}) to retrieve the top-$k$ results directly from a large document collection, substantially improving \textit{recall} over models that only re-rank the output of term-based retrieval.

As Figure~\ref{fig:tradeoff} illustrates, \sys{} can serve queries in tens or few hundreds of milliseconds. For instance, when used for re-ranking as in ``\sys{} (re-rank)'', it delivers over 170$\times$ speedup (and requires 14,000$\times$ fewer FLOPs) relative to existing BERT-based models, while being more effective than every non-BERT baseline (\secref{sec:evaluation:main-results} \& \ref{sec:evaluation:end-to-end}). \sys{}'s indexing---the only time it needs to feed documents through BERT---is also practical: it can index the MS MARCO collection of 9M passages in about 3 hours using a single server with four GPUs (\secref{sec:evaluation:indexing}), retaining its effectiveness with a space footprint of as little as few tens of GiBs. Our extensive ablation study (\secref{sec:evaluation:ablation}) shows that late interaction, its implementation via MaxSim operations, and crucial design choices within our BERT-based encoders are all essential to \sys{}'s effectiveness.

Our main contributions are as follows.
\begin{enumerate}
    \item  We propose \textit{late interaction} (\secref{sec:ColBERT:architecture}) as a paradigm for efficient and effective neural ranking. 
    
	\item We present \sys{} (\secref{sec:ColBERT:encoders} \& \ref{sec:ColBERT:training}), a highly-effective model that employs novel BERT-based query and document encoders within the late interaction paradigm.%
	
	\item We show how to leverage \sys{} both for re-ranking on top of a term-based retrieval model (\secref{sec:colbert:re-ranking}) and for searching a full collection using vector similarity indexes (\secref{sec:colbert:e2e-retrieval}). %
	
	\item We evaluate \sys{} on MS MARCO and TREC CAR, two recent passage search collections. %
	
\end{enumerate}

%% file: 2-related.tex
\section{Related Work}
\label{sec:related}

\textbf{Neural Matching Models.} Over the past few years, IR researchers have introduced numerous neural architectures for ranking. In this work, we compare against KNRM~\cite{xiong2017end, dai2018convolutional}, Duet~\cite{mitra2017learning, mitra2019updated}, ConvKNRM~\cite{dai2018convolutional}, and fastText+ConvKNRM~\cite{hofstatter2019effect}. KNRM proposes a differentiable kernel-pooling technique for extracting matching signals from an interaction matrix, while Duet combines signals from exact-match-based as well as embedding-based similarities for ranking. Introduced in 2018, ConvKNRM learns to match $n$-grams in the query and the document. Lastly, fastText+ConvKNRM (abbreviated fT+ConvKNRM) tackles the absence of rare words from typical word embeddings lists by adopting sub-word token embeddings.

In 2018, Zamani~\etal{}~\cite{zamani2018neural} introduced SNRM, a representation-focused IR model that encodes each query and each document as a single, sparse high-dimensional vector of ``latent terms''. By producing a sparse-vector representation for each document, SNRM is able to use a traditional IR inverted index for representing documents, allowing fast end-to-end retrieval. Despite highly promising results and insights, SNRM's effectiveness is substantially outperformed by the state of the art on the datasets with which it was evaluated (e.g., see \cite{yang2019critically, macavaney2019cedr}). While SNRM employs sparsity to allow using inverted indexes, we relax this assumption and compare a (dense) BERT-based representation-focused model against our late-interaction \sys{} in our ablation experiments in \secref{sec:evaluation:ablation}. For a detailed overview of existing neural ranking models, we refer the readers to two recent surveys of the literature \cite{mitra2018introduction, guo2019deep}.

\textbf{Language Model Pretraining for IR.} Recent work in NLU emphasizes the importance pre-training language representation models in an unsupervised fashion before subsequently fine-tuning them on downstream tasks. A notable example is BERT~\cite{devlin2018bert}, a bi-directional transformer-based language model whose fine-tuning advanced the state of the art on various NLU benchmarks. Nogueira~\etal{}~\cite{nogueira2019passage}, MacAvaney~\etal{}~\cite{macavaney2019cedr}, and Dai \& Callan~\cite{dai2019deeper} investigate incorporating such LMs (mainly BERT, but also ELMo~\cite{peters2018deep}) on different ranking datasets. As illustrated in Figure~\ref{fig:X} (c), the common approach (and the one adopted by Nogueira~\etal{} on MS MARCO and TREC CAR) is to feed the query--document pair through BERT and use an MLP on top of BERT's [CLS] output token to produce a relevance score. Subsequent work by Nogueira~\etal{}~\cite{nogueira2019multi} introduced duoBERT, which fine-tunes BERT to compare the relevance of a \textit{pair} of documents given a query. Relative to their single-document BERT, this gives duoBERT a 1\% MRR@10 advantage on MS MARCO while increasing the cost by at least 1.4$\times$.

\textbf{BERT Optimizations.} As discussed in \secref{sec:introduction}, these LM-based rankers can be highly expensive in practice. While ongoing efforts in the NLU literature for distilling \cite{jiao2019tinybert,tang2019distilling}, compressing \cite{zafrir2019q8bert}, and pruning \cite{michel2019sixteen} BERT can be instrumental in narrowing this gap, they generally achieve significantly smaller speedups than our re-designed architecture for IR, due to their generic nature, and more aggressive optimizations often come at the cost of lower quality. %

\textbf{Efficient NLU-based Models.} Recently, a direction emerged that employs expensive NLU computation offline. This includes doc2query~\cite{nogueira2019document} and DeepCT~\cite{dai2019context}. The doc2query model expands each document with a pre-defined number of synthetic queries queries generated by a seq2seq transformer model that is trained to generate \textit{queries} given a document. It then relies on a BM25 index for retrieval from the (expanded) documents. DeepCT uses BERT to produce the \textit{term frequency} component of BM25 in a context-aware manner, essentially representing a feasible realization of the term-independence assumption with neural networks \cite{mitra2019incorporating}. Lastly, docTTTTTquery~\cite{nogueiradoc2query} is identical to doc2query except that it fine-tunes a pre-trained model (namely, T5~\cite{raffel2019exploring}) for generating the predicted queries.

Concurrently with our drafting of this paper, Hofst\"{a}tter~\etal{}~\cite{hofstatter2019tu} published their Transformer-Kernel (TK) model. At a high level, TK improves the KNRM architecture described earlier: while KNRM employs kernel pooling on top of word-embedding-based interaction, TK uses a Transformer~\cite{vaswani2017attention} component for contextually encoding queries and documents before kernel pooling. TK establishes a new state-of-the-art for non-BERT models on MS MARCO (Dev); however, the best non-ensemble MRR@10 it achieves is 31\% while \sys{} reaches up to 36\%. Moreover, due to indexing document representations offline and employing a MaxSim-based late interaction mechanism, \sys{} is much more scalable, enabling end-to-end retrieval which is not supported by TK.

%% file: 3-ColBERT.tex
\begin{figure}[t]
\centering
\includegraphics[width=0.95\columnwidth]{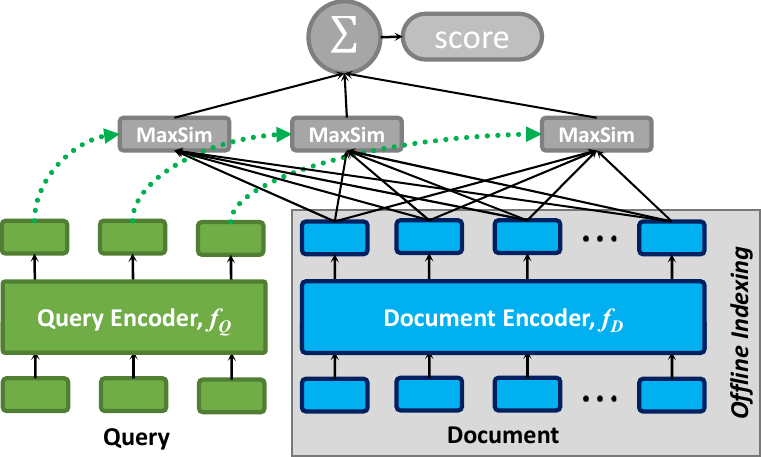}
\vspace*{0mm}
\caption{The general architecture of \sys{} given a query $q$ and a document $d$.}
\label{fig:ColBERT}
\vspace{-4mm}
\end{figure}

\section{\sys{}}
\label{sec:ColBERT}

\sys{} prescribes a simple framework for balancing the quality and cost of neural IR, particularly deep language models like BERT. As introduced earlier, delaying the query--document interaction can facilitate cheap neural re-ranking (i.e., through pre-computation) and even support practical end-to-end neural retrieval (i.e., through pruning via vector-similarity search). \sys{} addresses how to do so while still preserving the effectiveness of state-of-the-art models, which condition the bulk of their computations on the joint query--document pair.

Even though \sys{}'s late-interaction framework can be applied to a wide variety of architectures (e.g., CNNs, RNNs, transformers, etc.), we choose to focus this work on bi-directional transformer-based encoders (i.e., BERT) owing to their state-of-the-art effectiveness yet very high computational cost.

\subsection{Architecture}
\label{sec:ColBERT:architecture}

Figure~\ref{fig:ColBERT} depicts the general architecture of \sys{}, which comprises: (a) a query encoder $f_Q$, (b) a document encoder $f_D$, and (c) the late interaction mechanism. Given a query $q$ and document $d$, $f_Q$ encodes $q$ into a bag of fixed-size embeddings $E_q$ while $f_D$ encodes $d$ into another bag $E_d$. Crucially, each embeddings in $E_q$ and $E_d$ is \textit{contextualized} based on the other terms in $q$ or $d$, respectively. We describe our BERT-based encoders in \secref{sec:ColBERT:encoders}.

Using $E_q$ and $E_d$, \sys{} computes the relevance score between $q$ and $d$ via late interaction, which we define as a summation of maximum similarity (MaxSim) operators. In particular, we find the maximum cosine similarity of each $v \in E_q$ with vectors in $E_d$, and combine the outputs via summation. Besides cosine, we also evaluate squared L2 distance as a measure of vector similarity. Intuitively, this interaction mechanism softly \textit{searches} for each query term $t_q$---in a manner that reflects its context in the query---against the document's embeddings, quantifying the strength of the ``match'' via the largest similarity score between $t_q$ and a document term $t_d$. Given these term scores, it then estimates the document relevance by summing the matching evidence across all query terms.

While more sophisticated matching is possible with other choices such as deep convolution and attention layers (i.e., as in typical interaction-focused models), a summation of maximum similarity computations has two distinctive characteristics. First, it stands out as a particularly cheap interaction mechanism, as we examine its FLOPs in \secref{sec:evaluation:main-results}. Second, and more importantly, it is amenable to highly-efficient pruning for top-$k$ retrieval, as we evaluate in \secref{sec:evaluation:end-to-end}. This enables using vector-similarity algorithms for skipping documents without materializing the full interaction matrix or even considering each document in isolation. Other cheap choices (e.g., a summation of \textit{average} similarity scores, instead of maximum) are possible; however, many are less amenable to pruning. In \secref{sec:evaluation:ablation}, we conduct an extensive ablation study that empirically verifies the advantage of our MaxSim-based late interaction against alternatives.

\subsection{Query \& Document Encoders}
\label{sec:ColBERT:encoders}

Prior to late interaction, \sys{} encodes each query or document into a bag of embeddings, employing BERT-based encoders. We share a single BERT model among our query and document encoders but distinguish input sequences that correspond to queries and documents by prepending a special token \texttt{[Q]} to queries and another token \texttt{[D]} to documents. %

\textbf{Query Encoder.} Given a textual query $q$, we tokenize it into its BERT-based WordPiece~\cite{wu2016google} tokens $q_1 q_2 ... q_l$. We prepend the token \texttt{[Q]} to the query. We place this token right after BERT's sequence-start token \texttt{[CLS]}. If the query has fewer than a pre-defined number of tokens $N_q$, we pad it with BERT's special \texttt{[mask]} tokens up to length $N_q$ (otherwise, we truncate it to the first $N_q$ tokens). This padded sequence of input tokens is then passed into BERT's deep transformer architecture, which computes a contextualized representation of each token.

We denote the padding with masked tokens as \textbf{query augmentation}, a step that allows BERT to produce query-based embeddings at the positions corresponding to these masks. Query augmentation is intended to serve as a soft, differentiable mechanism for learning to expand queries with new terms or to re-weigh existing terms based on their importance for matching the query. As we show in \secref{sec:evaluation:ablation}, this operation is essential for \sys{}'s effectiveness.

Given BERT's representation of each token, our encoder passes the contextualized output representations through a linear layer with no activations. This layer serves to control the dimension of \sys{}'s embeddings, producing $m$-dimensional embeddings for the layer's output size $m$. As we discuss later in more detail, we typically fix $m$ to be much smaller than BERT's fixed hidden dimension. %

While \sys{}'s embedding dimension has limited impact on the efficiency of query encoding, this step is crucial for controlling the space footprint of documents, as we show in \secref{sec:evaluation:indexing}. In addition, it can have a significant impact on query execution time, particularly the time taken for transferring the document representations onto the GPU from system memory (where they reside before processing a query). In fact, as we show in \secref{sec:evaluation:main-results}, gathering, stacking, and transferring the embeddings from CPU to GPU can be the most expensive step in re-ranking with \sys{}. Finally, the output embeddings are normalized so each has L2 norm equal to one. The result is that the dot-product of any two embeddings becomes equivalent to their cosine similarity, falling in the $[-1, 1]$ range.

\textbf{Document Encoder.} Our document encoder has a very similar architecture. We first segment a document $d$ into its constituent tokens $d_1 d_2 ... d_m$, to which we prepend BERT's start token \texttt{[CLS]} followed by our special token \texttt{[D]} that indicates a document sequence. Unlike queries, we do not append \texttt{[mask]} tokens to documents. After passing this input sequence through BERT and the subsequent linear layer, the document encoder filters out the embeddings corresponding to punctuation symbols, determined via a pre-defined list. This filtering is meant to reduce the number of embeddings per document, as we hypothesize that (even contextualized) embeddings of punctuation are unnecessary for effectiveness. %

In summary, given $q = q_0 q_1 ... q_l$ and $d = d_0 d_1 ... d_n$, we compute the bags of embeddings $E_q$ and $E_d$ in the following manner, where $\#$ refers to the \texttt{[mask]} tokens:
\begin{align}
\label{equation:encoders}
    E_q & := \texttt{Normalize}(\; \texttt{CNN}(\; \texttt{BERT}(``[Q] q_0 q_1 ... q_l \# \# ... \#") \;) \;) \\
    E_d & := \texttt{Filter}(\; \texttt{Normalize}(\; \texttt{CNN}(\; \texttt{BERT}(``[D] d_0 d_1 ... d_n") \;) \;)  \;)
\end{align}

\subsection{Late Interaction}

Given the representation of a query $q$ and a document $d$, the relevance score of $d$ to $q$, denoted as $S_{q,d}$, is estimated via late interaction between their bags of contextualized embeddings. As mentioned before, this is conducted as a sum of maximum similarity computations, namely cosine similarity (implemented as dot-products due to the embedding normalization) or squared L2 distance. %

\begin{align}
    S_{q,d} & := \sum_{i \in [|E_q|]} \max_{j \in [|E_d|]} E_{q_i} \cdot E_{d_j}^{T}
\end{align}

\label{sec:ColBERT:training}

\sys{} is differentiable end-to-end. We fine-tune the BERT encoders and train from scratch the additional parameters (i.e., the linear layer and the [Q] and [D] markers' embeddings) using the Adam~\cite{kingma2014adam} optimizer. Notice that our interaction mechanism has no trainable parameters. Given a triple $\langle q, d^+, d^- \rangle$ with query $q$, positive document $d^+$ and negative document $d^-$, \sys{} is used to produce a score for each document individually and is optimized via pairwise softmax cross-entropy loss over the computed scores of $d^+$ and $d^-$.

\subsection{Offline Indexing: Computing \& Storing Document Embeddings}
\label{sec:ColBERT:indexing}

By design, \sys{} isolates almost all of the computations between queries and documents, largely to enable pre-computing document representations offline. At a high level, our indexing procedure is straight-forward: we proceed over the documents in the collection in batches, running our document encoder $f_D$ on each batch and storing the output embeddings per document. Although indexing a set of documents is an offline process, we incorporate a few simple optimizations for enhancing the throughput of indexing. As we show in \secref{sec:evaluation:indexing}, these optimizations can considerably reduce the offline cost of indexing.

To begin with, we exploit multiple GPUs, if available, for faster encoding of batches of documents in parallel. When batching, we pad all documents to the maximum length of a document \textit{within} the batch.\footnote{The public BERT implementations we saw simply pad to a pre-defined length.} To make capping the sequence length on a per-batch basis more effective, our indexer proceeds through documents in groups of $B$ (e.g., $B=$~100,000) documents. It sorts these documents by length and then feeds batches of $b$ (e.g., $b=$~128) documents of comparable length through our encoder. This length-based bucketing is sometimes refered to as a \texttt{BucketIterator} in some libraries (e.g., allenNLP). Lastly, while most computations occur on the GPU, we found that a non-trivial portion of the indexing time is spent on pre-processing the text sequences, primarily BERT's WordPiece tokenization. Exploiting that these operations are independent across documents in a batch, we parallelize the pre-processing across the available CPU cores.

Once the document representations are produced, they are saved to disk using 32-bit or 16-bit values to represent each dimension. As we describe in \secref{sec:colbert:re-ranking} and \ref{sec:colbert:e2e-retrieval}, these representations are either simply loaded from disk for ranking or are subsequently indexed for vector-similarity search, respectively.  %

\subsection{Top-$k$ Re-ranking with \sys{}}
\label{sec:colbert:re-ranking}

Recall that \sys{} can be used for re-ranking the output of another retrieval model, typically a term-based model, or directly for end-to-end retrieval from a document collection. In this section, we discuss how we use \sys{} for ranking a small set of $k$ (e.g., $k=1000$) documents given a query $q$. Since $k$ is small, we rely on batch computations to exhaustively score each document (unlike our approach in \secref{sec:colbert:e2e-retrieval}). To begin with, our query serving sub-system loads the indexed documents representations into memory, representing each document as a matrix of embeddings.

Given a query $q$, we compute its bag of contextualized embeddings $E_q$ (Equation~\ref{equation:encoders}) and, concurrently, gather the document representations into a 3-dimensional tensor $D$ consisting of $k$ document matrices. We pad the $k$ documents to their maximum length to facilitate batched operations, and move the tensor $D$ to the GPU's memory. On the GPU, we compute a batch dot-product of $E_q$ and $D$, possibly over multiple mini-batches. The output materializes a 3-dimensional tensor that is a collection of cross-match matrices between $q$ and each document. To compute the score of each document, we reduce its matrix across document terms via a max-pool (i.e., representing an exhaustive implementation of our MaxSim computation) and reduce across query terms via a summation. Finally, we sort the $k$ documents by their total scores.

Relative to existing neural rankers (especially, but not exclusively, BERT-based ones), this computation is very cheap that, in fact, its cost is dominated by the cost of gathering and transferring the pre-computed embeddings. To illustrate, ranking $k$ documents via typical BERT rankers requires feeding BERT $k$ different inputs each of length $l = |q| + |d_i|$ for query $q$ and documents $d_i$, where attention has quadratic cost in the length of the sequence. In contrast, \sys{} feeds BERT only a single, much shorter sequence of length $l=|q|$. Consequently, \sys{} is not only cheaper, it also scales much better with $k$ as we examine in \secref{sec:evaluation:main-results}.

\subsection{End-to-end Top-$k$ Retrieval with \sys{}}
\label{sec:colbert:e2e-retrieval}

As mentioned before, \sys{}'s late-interaction operator is specifically designed to enable end-to-end retrieval from a large collection, largely to improve recall relative to term-based retrieval approaches. This section is concerned with cases where the number of documents to be ranked is too large for exhaustive evaluation of each possible candidate document, particularly when we are only interested in the highest scoring ones. Concretely, we focus here on retrieving the top-$k$ results directly from a large document collection with $N$ (e.g., $N=10,000,000$) documents, where $k \ll N$.   

To do so, we leverage the pruning-friendly nature of the MaxSim operations at the backbone of late interaction. Instead of applying MaxSim between one of the query embeddings and all of one document's embeddings, we can use fast vector-similarity data structures to efficiently conduct this search between the query embedding and \textit{all} document embeddings across the full collection. For this, we employ an off-the-shelf library for large-scale vector-similarity search, namely \texttt{faiss}~\cite{JDH17} from Facebook.\footnote{https://github.com/facebookresearch/faiss}%
In particular, at the end of offline indexing (\secref{sec:ColBERT:indexing}), we maintain a mapping from each embedding to its document of origin and then index all document embeddings into \texttt{faiss}.

Subsequently, when serving queries, we use a two-stage procedure to retrieve the top-$k$ documents from the entire collection. Both stages rely on \sys{}'s scoring: the first is an approximate stage aimed at filtering while the second is a refinement stage. For the first stage, we concurrently issue $N_q$ vector-similarity queries (corresponding to each of the embeddings in $E_q$) onto our \texttt{faiss} index. This retrieves the top-$k'$ (e.g., $k'=k/2$) matches for that vector over all document embeddings. We map each of those to its document of origin, producing $N_q \times k'$ document IDs, only $K \leq N_q \times k'$ of which are unique. These $K$ documents likely contain one or more embeddings that are highly similar to the query embeddings. For the second stage, we refine this set by exhaustively re-ranking \textit{only} those $K$ documents in the usual manner described in \secref{sec:colbert:re-ranking}.

In our \texttt{faiss}-based implementation, we use an \texttt{IVFPQ} index (``inverted file with product quantization''). This index partitions the embedding space into $P$ (e.g., $P=1000$) cells based on $k$-means clustering and then assigns each document embedding to its nearest cell based on the selected vector-similarity metric. For serving queries, when searching for the top-$k'$ matches for a single query embedding, only the nearest $p$ (e.g., $p=10$) partitions are searched. To improve memory efficiency, every embedding is divided into $s$ (e.g., $s=16$) sub-vectors, each represented using one byte. Moreover, the index conducts the similarity computations in this compressed domain, leading to cheaper computations and thus faster search.

%% file: 4-experiments.tex
\section{Experimental Evaluation}
\label{sec:evaluation}

We now turn our attention to empirically testing \sys{}, addressing the following research questions.

\textbf{RQ$_1$}: In a typical re-ranking setup, how well can \sys{} bridge the existing gap (highlighted in \secref{sec:introduction}) between highly-efficient and highly-effective neural models? (\secref{sec:evaluation:main-results})

\textbf{RQ$_2$}: Beyond re-ranking, can \sys{} effectively support end-to-end retrieval directly from a large collection? (\secref{sec:evaluation:end-to-end})

\textbf{RQ$_3$}: What does each component of \sys{} (e.g., late interaction, query augmentation) contribute to its quality? (\secref{sec:evaluation:ablation})

\textbf{RQ$_4$}: What are \sys{}'s indexing-related costs in terms of offline computation and memory overhead? (\secref{sec:evaluation:indexing})

\begin{table*}[t]
\small
\centering
\begin{tabularx}{\textwidth}{@{}lccccc@{}}
\toprule
\textbf{Method} & \hspace{14mm} & \textbf{MRR@10 (Dev)} & \textbf{MRR@10 (Eval)} & \textbf{Re-ranking Latency (ms)} & \textbf{FLOPs/query} \\ \toprule

BM25 (official)      &&    16.7    & 16.5 &     -   & -     \\
\midrule
\texttt{KNRM}       &&    19.8    & 19.8 &  3     &   592M  (0.085$\times$)  \\
\texttt{Duet}       &&    24.3    & 24.5 & 22   &     159B  (23$\times$)  \\
\texttt{fastText+ConvKNRM}       && 29.0  &  27.7  & 28  &  78B  (11$\times$)  \\
\BertBase{} \cite{nogueira2019passage}      &&   34.7   & -  &   10,700   &   97T  (13,900$\times$)  \\
\BertBase{} (our training)      &&   36.0 &  -   &   10,700  &  97T (13,900$\times$) \\
\BertLarge{} \cite{nogueira2019passage}      &&     36.5   & 35.9 &   32,900     &    340T (48,600$\times$)  \\
\midrule
ColBERT  (over \BertBase{})  &&    34.9   & 34.9 & 61    & 7B   (1$\times$)      \\

\bottomrule
\end{tabularx}
\vspace{2mm}
\caption{``Re-ranking'' results on MS MARCO. Each neural model re-ranks the official top-1000 results produced by BM25. Latency is reported for re-ranking only. To obtain the end-to-end latency in Figure~\ref{fig:tradeoff}, we add the BM25 latency from Table~\ref{table:msmarco-fullranking}.}
\label{table:msmarco-reranking}
\vspace{-3mm}
\end{table*}

\begin{table*}[t]
\small
\centering
\begin{tabularx}{\textwidth}{@{}lccccccc@{}}
\toprule
\textbf{Method} & \hspace{5mm} & \textbf{MRR@10 (Dev)} & \textbf{MRR@10 (Local Eval)} & \textbf{Latency (ms)} & \textbf{Recall@50} & \textbf{Recall@200} & \textbf{Recall@1000} \\ \toprule
BM25 (official)      &&    16.7    & -  &   -    &  - &  - & 81.4   \\
BM25 (Anserini)      &&     18.7   & 19.5  &  62     & 59.2 & 73.8 & 85.7  \\
\texttt{doc2query}       &&   21.5    & 22.8 & 85   & 64.4 & 77.9 & 89.1 \\
\texttt{DeepCT}       &&     24.3   & - &  62 \textit{(est.)}    & 69 \cite{dai2019context} & 82 \cite{dai2019context} & 91 \cite{dai2019context}  \\
\texttt{docTTTTTquery}       &&   27.7  & 28.4 &  87     & 75.6 & 86.9 & 94.7 \\
\midrule
ColBERT$_\textnormal{L2}$  (re-rank)     &&     34.8  & 36.4 &  -       & 75.3 & 80.5 & 81.4  \\
ColBERT$_\textnormal{L2}$  (end-to-end)     &&     36.0  & 36.7 &  458       & 82.9 & 92.3 & 96.8  \\
\bottomrule
\end{tabularx}
\vspace{2mm}
\caption{End-to-end retrieval results on MS MARCO. Each model retrieves the top-1000 documents per query \textit{directly} from the entire 8.8M document collection.}
\vspace{-5mm}
\label{table:msmarco-fullranking}
\end{table*}

\subsection{Methodology}
\label{sec:evaluation:methodology}

\subsubsection{Datasets \& Metrics} Similar to related work \cite{nogueira2019document,dai2019context,nogueira2019multi}, we conduct our experiments on the MS MARCO Ranking \cite{nguyen2016ms} (henceforth, MS MARCO) and TREC Complex Answer Retrieval (TREC-CAR) \cite{dietz2017trec} datasets. Both of these recent datasets provide large training data of the scale that facilitates training and evaluating deep neural networks. We describe both in detail below.

\textbf{MS MARCO.} MS MARCO is a dataset (and a corresponding competition) introduced by Microsoft in 2016 for reading comprehension and adapted in 2018 for retrieval. It is a collection of 8.8M passages from Web pages, which were gathered from Bing's results to 1M real-world queries. Each query is associated with \textit{sparse} relevance judgements of one (or very few) documents marked as relevant and no documents explicitly indicated as irrelevant. Per the official evaluation, we use MRR@10 to measure effectiveness. %

We use three sets of queries for evaluation. The official development and evaluation sets contain roughly 7k queries. However, the relevance judgements of the evaluation set are held-out by Microsoft and effectiveness results can only be obtained by submitting to the competition's organizers. We submitted our main re-ranking \sys{} model for the results in \secref{sec:evaluation:main-results}. In addition, the collection includes roughly 55k queries (with labels) that are provided as additional validation data. We re-purpose a random sample of 5k queries among those (i.e., ones not in our development or training sets) as a ``local'' evaluation set. Along with the official development set, we use this held-out set for testing our models as well as baselines in \secref{sec:evaluation:end-to-end}. We do so to avoid submitting multiple variants of the same model at once, as the organizers discourage too many submissions by the same team.

\textbf{TREC CAR.} Introduced by Dietz~\cite{dietz2017trec}~\etal{} in 2017, TREC CAR is a synthetic dataset based on Wikipedia that consists of about 29M passages. Similar to related work~\cite{nogueira2019passage}, we use the first four of five pre-defined folds for training and the fifth for validation. This amounts to roughly 3M queries generated by concatenating the title of a Wikipedia page with the heading of one of its sections. That section's passages are marked as relevant to the corresponding query. Our evaluation is conducted on the test set used in TREC 2017 CAR, which contains 2,254 queries.

\subsubsection{Implementation} Our \sys{} models are implemented using Python 3 and PyTorch 1. We use the popular \texttt{transformers}\footnote{https://github.com/huggingface/transformers} library for the pre-trained BERT model. Similar to \cite{nogueira2019passage}, we fine-tune all \sys{} models with learning rate $3\times10^{-6}$ with a batch size 32. We fix the number of embeddings per query at $N_q=32$. We set our \sys{} embedding dimension $m$ to be 128; \secref{sec:evaluation:indexing} demonstrates \sys{}'s robustness to a wide range of embedding dimensions.

For MS MARCO, we initialize the BERT components of the \sys{} query and document encoders using Google's official pre-trained \BertBase{} model. Further, we train all models for 200k iterations. For TREC CAR, we follow related work \cite{nogueira2019passage,dai2019context} and use a different pre-trained model to the official ones. To explain, the official BERT models were pre-trained on Wikipedia, which is the source of TREC CAR's training and test sets. To avoid leaking test data into train, Nogueira and Cho's~\cite{nogueira2019passage} pre-train a randomly-initialized BERT model on the Wiki pages corresponding to training subset of TREC CAR. They release their \BertLarge{} pre-trained model, which we fine-tune for \sys{}'s experiments on TREC CAR. Since fine-tuning this model is significantly slower than \BertBase{}, we train on TREC CAR for only 125k iterations.

In our re-ranking results, unless stated otherwise, we use 4 bytes per dimension in our embeddings and employ cosine as our vector-similarity function. For end-to-end ranking, we use (squared) L2 distance, as we found our \texttt{faiss} index was faster at L2-based retrieval. For our \texttt{faiss} index, we set the number of partitions to $P=$2,000, and search the nearest $p=10$ to each query embedding to retrieve $k'=k=1000$ document vectors per query embedding. We divide each embedding into $s=16$ sub-vectors, each encoded using one byte. To represent the index used for the second stage of our end-to-end retrieval procedure, we use 16-bit values per dimension.

\subsubsection{Hardware \& Time Measurements} To evaluate the latency of neural re-ranking models in \secref{sec:evaluation:main-results}, we use a single Tesla V100 GPU that has 32 GiBs of memory on a server with two Intel Xeon Gold 6132 CPUs, each with 14 physical cores (24 hyperthreads), and 469 GiBs of RAM. For the mostly CPU-based retrieval experiments in \secref{sec:evaluation:end-to-end} and the indexing experiments in \secref{sec:evaluation:indexing}, we use another server with the same CPU and system memory specifications but which has four Titan V GPUs attached, each with 12 GiBs of memory. Across all experiments, only one GPU is dedicated per query for retrieval (i.e., for methods with neural computations) but we use up to all four GPUs during indexing.

\subsection{Quality--Cost Tradeoff: Top-$k$ Re-ranking}
\label{sec:evaluation:main-results}

In this section, we examine \sys{}'s efficiency and effectiveness at re-ranking the top-$k$ results extracted by a bag-of-words retrieval model, which is the most typical setting for testing and deploying neural ranking models. We begin with the MS MARCO dataset. We compare against KNRM, Duet, and fastText+ConvKNRM, a representative set of neural matching models that have been previously tested on MS MARCO. In addition, we compare against the natural adaptation of BERT for ranking by Nogueira and Cho~\cite{nogueira2019passage}, in particular, \BertBase{} and its deeper counterpart \BertLarge{}. We also report results for ``\BertBase{} (our training)'', which is based on Nogueira and Cho's base model (including hyperparameters) but is trained with the same loss function as \sys{} (\secref{sec:ColBERT:training}) for 200k iterations, allowing for a more direct comparison of the results.

We report the competition's official metric, namely MRR@10, on the validation set (Dev) and the evaluation set (Eval). We also report the re-ranking latency, which we measure using a single Tesla V100 GPU, and the FLOPs per query for each neural ranking model. For \sys{}, our reported latency subsumes the entire computation from gathering the document representations, moving them to the GPU, tokenizing then encoding the query, and applying late interaction to compute document scores. For the baselines, we measure the scoring computations on the GPU and exclude the CPU-based text preprocessing (similar to \cite{hofstatter2019let}). In principle, the baselines can pre-compute the majority of this preprocessing (e.g., document tokenization) offline and parallelize the rest across documents online, leaving only a negligible cost. We estimate the FLOPs per query of each model using the torchprofile\footnote{https://github.com/mit-han-lab/torchprofile} library.

We now proceed to study the results, which are reported in Table~\ref{table:msmarco-reranking}. To begin with, we notice the fast progress from KNRM in 2017 to the BERT-based models in 2019, manifesting itself in over 16\% increase in MRR@10. As described in \secref{sec:introduction}, the simultaneous increase in computational cost is difficult to miss. Judging by their rather monotonic pattern of increasingly larger cost and higher effectiveness, these results appear to paint a picture where expensive models are necessary for high-quality ranking.

In contrast with this trend, \sys{} (which employs late interaction over \BertBase{}) performs no worse than the original adaptation of \BertBase{} for ranking by Nogueira and Cho~\cite{nogueira2019passage,nogueira2019multi} and is only marginally less effective than \BertLarge{} and our training of \BertBase{} (described above). While highly competitive in effectiveness, \sys{} is orders of magnitude cheaper than \BertBase{}, in particular, by over 170$\times$ in latency and 13,900$\times$ in FLOPs. This highlights the expressiveness of our proposed late interaction mechanism, particularly when coupled with a powerful pre-trained LM like BERT. While \sys{}'s re-ranking latency is slightly higher than the non-BERT re-ranking models shown (i.e., by 10s of milliseconds), this difference is explained by the time it takes to gather, stack, and transfer the document embeddings to the GPU. In particular, the query encoding and interaction in \sys{} consume only 13 milliseconds of its total execution time. We note that \sys{}'s latency and FLOPs can be considerably reduced by padding queries to a shorter length, using smaller vector dimensions (the MRR@10 of which is tested in \secref{sec:evaluation:indexing}), employing quantization of the document vectors, and storing the embeddings on GPU if sufficient memory exists. We leave these directions for future work.

\begin{figure}[h]
\centering
\includegraphics[width=\columnwidth]{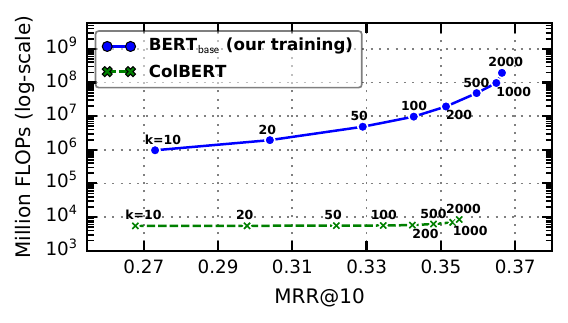}
\vspace{-7mm}
\caption{FLOPs (in millions) and MRR@10 as functions of the re-ranking depth $k$. Since the official BM25 ranking is not ordered, the initial top-$k$ retrieval is conducted with Anserini's BM25.}
\label{fig:value-of-k}
\vspace{-3mm}
\end{figure}

Diving deeper into the quality--cost tradeoff between BERT and \sys{}, Figure~\ref{fig:value-of-k} demonstrates the relationships between FLOPs and effectiveness (MRR@10) as a function of the re-ranking depth $k$ when re-ranking the top-$k$ results by BM25, comparing \sys{} and \BertBase{} (our training). We conduct this experiment on MS MARCO (Dev). We note here that as the official top-1000 ranking does not provide the BM25 order (and also lacks documents beyond the top-1000 per query), the models in this experiment re-rank the Anserini~\cite{yang2018anserini} toolkit's BM25 output. Consequently, both MRR@10 values at $k=1000$ are slightly higher from those reported in Table~\ref{table:msmarco-reranking}.

Studying the results in Figure~\ref{fig:value-of-k}, we notice that not only is \sys{} much cheaper than BERT for the same model size (i.e., 12-layer ``base'' transformer encoder), it also scales better with the number of ranked documents. In part, this is because \sys{} only needs to process the query once, irrespective of the number of documents evaluated. For instance, at $k=10$, BERT requires nearly 180$\times$ more FLOPs than \sys{}; at $k=1000$, BERT's overhead jumps to 13,900$\times$. It then reaches 23,000$\times$ at $k=2000$. In fact, our informal experimentation shows that this orders-of-magnitude gap in FLOPs makes it practical to run \sys{} entirely on the CPU, although CPU-based re-ranking lies outside our scope.

\begin{table}[h]
\small
\centering
\begin{tabularx}{0.75\columnwidth}{@{}Xllll@{}}
\toprule
Method & MAP & MRR@10   \\ \midrule
BM25 (Anserini)     & 15.3 &      -                 \\
doc2query      &   18.1  &       -          \\
DeepCT      &  24.6  &  33.2                    \\
\midrule
BM25 + \BertBase{}          &  31.0 &           -     \\
BM25 + \BertLarge{}          &  33.5   &          -       \\
\midrule
BM25 + ColBERT       & 31.3  &  44.3        \\
\bottomrule
\end{tabularx}
\caption{Results on TREC CAR.}
\label{table:treccar-results}
\vspace{-2mm}
\end{table}

Having studied our results on MS MARCO, we now consider TREC CAR, whose official metric is MAP. Results are summarized in Table~\ref{table:treccar-results}, which includes a number of important baselines (BM25, doc2query, and DeepCT) in addition to re-ranking baselines that have been tested on this dataset. These results directly mirror those with MS MARCO.

\subsection{End-to-end Top-$k$ Retrieval}
\label{sec:evaluation:end-to-end}

Beyond cheap re-ranking, \sys{} is amenable to top-$k$ retrieval directly from a full collection. Table~\ref{table:msmarco-fullranking} considers full retrieval, wherein each model retrieves the top-1000 documents directly from MS MARCO's 8.8M documents per query. In addition to MRR@10 and latency in milliseconds, the table reports Recall@50, Recall@200, and Recall@1000, important metrics for a full-retrieval model that essentially filters down a large collection on a per-query basis.

We compare against BM25, in particular MS MARCO's official BM25 ranking as well as a well-tuned baseline based on the Anserini toolkit.\footnote{http://anserini.io/} While many other traditional models exist, we are not aware of any that substantially outperform Anserini's BM25 implementation (e.g., see RM3 in \cite{nogueira2019document}, LMDir in \cite{dai2019context}, or Microsoft's proprietary feature-based RankSVM on the leaderboard).

We also compare against doc2query, DeepCT, and docTTTTTquery. All three rely on a traditional bag-of-words model (primarily BM25) for retrieval. Crucially, however, they re-weigh the frequency of terms per document and/or expand the set of terms in each document before building the BM25 index. In particular, doc2query expands each document with a pre-defined number of synthetic queries generated by a seq2seq transformer model (which docTTTTquery replaced with a pre-trained language model, T5~\cite{raffel2019exploring}). In contrast, DeepCT uses BERT to produce the term frequency component of BM25 in a context-aware manner.

For the latency of Anserini's BM25, doc2query, and docTTTTquery, we use the authors' \cite{nogueira2019document,nogueiradoc2query} Anserini-based implementation. While this implementation supports multi-threading, it only utilizes parallelism across different queries. We thus report single-threaded latency for these models, noting that simply parallelizing their computation over \textit{shards} of the index can substantially decrease their already-low latency. For DeepCT, we only estimate its latency using that of BM25 (as denoted by \textit{(est.)} in the table), since DeepCT re-weighs BM25's term frequency without modifying the index otherwise.\footnote{In practice, a myriad of reasons could still cause DeepCT's latency to differ slightly from BM25's. For instance, the top-$k$ pruning strategy employed, if any, could interact differently with a changed distribution of scores.} As discussed in \secref{sec:evaluation:methodology}, we use \sys{}$_\textnormal{L2}$ for end-to-end retrieval, which employs negative squared L2 distance as its vector-similarity function. For its latency, we measure the time for \texttt{faiss}-based candidate filtering and the subsequent re-ranking. In this experiment, \texttt{faiss} uses all available CPU cores.

Looking at Table~\ref{table:msmarco-fullranking}, we first see Anserini's BM25 baseline at 18.7 MRR@10, noticing its very low latency as implemented in Anserini (which extends the well-known Lucene system), owing to both very cheap operations and decades of bag-of-words top-$k$ retrieval optimizations. The three subsequent baselines, namely doc2query, DeepCT, and docTTTTquery, each brings a decisive enhancement to effectiveness. These improvements come at negligible overheads in latency, since these baselines ultimately rely on BM25-based retrieval. The most effective among these three, docTTTTquery, demonstrates a massive 9\% gain over vanilla BM25 by fine-tuning the recent language model T5.

Shifting our attention to \sys{}'s end-to-end retrieval effectiveness, we see its major gains in MRR@10 over all of these end-to-end models. In fact, using \sys{} in the end-to-end setup is superior in terms of MRR@10 to re-ranking with the same model due to the improved recall. Moving beyond MRR@10, we also see large gains in Recall@$k$ for $k$ equals to 50, 200, and 1000. For instance, its Recall@50 actually exceeds the official BM25's Recall@1000 and even all but docTTTTTquery's Recall@200, emphasizing the value of end-to-end retrieval (instead of just re-ranking) with \sys{}.

\subsection{Ablation Studies}
\label{sec:evaluation:ablation}

\begin{figure}[h]
\centering
\includegraphics[width=\columnwidth]{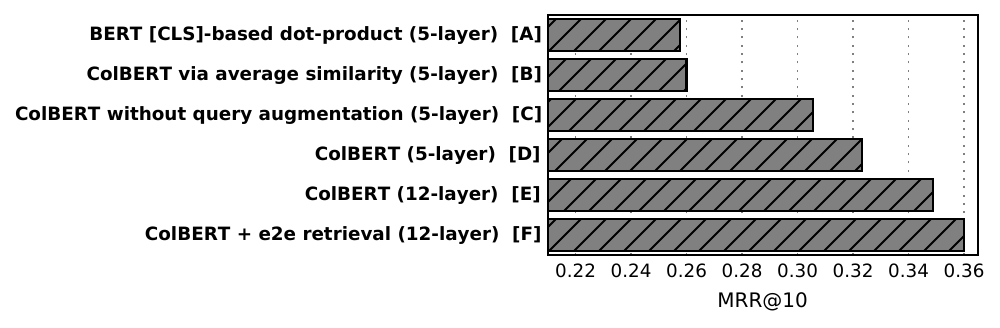}
\vspace{-7mm}
\caption{Ablation results on MS MARCO (Dev). Between brackets is the number of BERT layers used in each model.}
\label{fig:ablation}
\vspace{-3mm}
\end{figure}

The results from \secref{sec:evaluation:main-results} indicate that \sys{} is highly effective despite the low cost and simplicity of its late interaction mechanism. To better understand the source of this effectiveness, we examine a number of important details in \sys{}'s interaction and encoder architecture. For this ablation, we report MRR@10 on the validation set of MS MARCO in Figure~\ref{fig:ablation}, which shows our main \textit{re-ranking} \sys{} model [E], with MRR@10 of 34.9\%.

Due to the cost of training all models, we train a copy of our main model that retains only the first 5 layers of BERT out of 12 (i.e., model [D]) and similarly train all our ablation models for 200k iterations with five BERT layers. To begin with, we ask if the fine-granular \textit{interaction} in late interaction is necessary. Model [A] tackles this question: it uses BERT to produce a single embedding vector for the query and another for the document, extracted from BERT's [CLS] contextualized embedding and expanded through a linear layer to dimension 4096 (which equals $N_q \times 128 = 32\times128$). Relevance is estimated as the inner product of the query's and the document's embeddings, which we found to perform better than cosine similarity for single-vector re-ranking. As the results show, this model is considerably less effective than \sys{}, reinforcing the importance of late interaction.

Subsequently, we ask if our MaxSim-based late interaction is better than other simple alternatives. We test a model [B] that replaces \sys{}'s maximum similarity with \textit{average} similarity. The results suggest the importance of individual terms in the query paying special attention to particular terms in the document. Similarly, the figure emphasizes the importance of our query augmentation mechanism: without query augmentation [C], \sys{} has a noticeably lower MRR@10. Lastly, we see the impact of end-to-end retrieval not only on recall but also on MRR@10. By retrieving directly from the full collection, \sys{} is able to retrieve to the top-10 documents missed entirely from BM25's top-1000.

\subsection{Indexing Throughput \& Footprint}
\label{sec:evaluation:indexing}

\begin{figure}[h]
\centering
\includegraphics[width=\columnwidth]{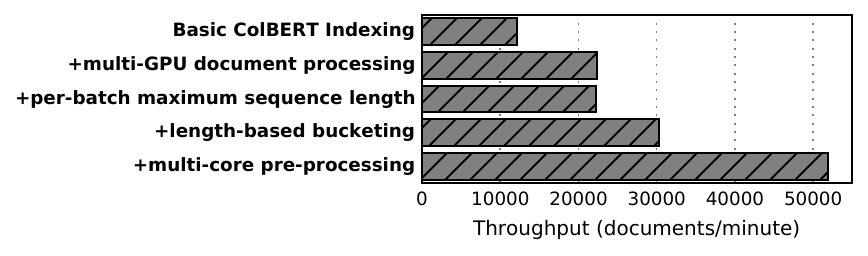}
\vspace{-7mm}
\caption{Effect of \sys{}'s indexing optimizations on the offline indexing throughput.}
\label{fig:throughput}
\end{figure}

Lastly, we examine the indexing throughput and space footprint of \sys{}. Figure~\ref{fig:throughput} reports indexing throughput on MS MARCO documents with \sys{} and four other ablation settings, which individually enable optimizations described in \secref{sec:ColBERT:indexing} on top of basic batched indexing. Based on these throughputs, \sys{} can index MS MARCO in about three hours. Note that any BERT-based model must incur the computational cost of processing each document at least once. While \sys{} encodes each document with BERT exactly once, existing BERT-based rankers would repeat similar computations on possibly hundreds of documents for each query.

\begin{table}[h]
\small
\centering
\begin{tabularx}{\columnwidth}{@{}Xrrrr@{}}
\toprule
Setting & Dimension($m$) & Bytes/Dim & Space(GiBs) & MRR@10   \\ \midrule
Re-rank Cosine & 128 & 4 & 286 & 34.9 \\
End-to-end L2 & 128 & 2 & 154 & 36.0 \\
Re-rank L2 & 128 & 2 & 143 & 34.8 \\
Re-rank Cosine & 48 & 4 & 54 & 34.4 \\
Re-rank Cosine & 24 & 2 & 27 & 33.9 \\
\bottomrule
\end{tabularx}
\caption{Space Footprint vs MRR@10 (Dev) on MS MARCO.}
\label{table:footprint}
\vspace{-4mm}
\end{table}

Table~\ref{table:footprint} reports the space footprint of \sys{} under various settings as we reduce the embeddings dimension and/or the bytes per dimension. Interestingly, the most space-efficient setting, that is, re-ranking with cosine similarity with 24-dimensional vectors stored as 2-byte floats, is only 1\% worse in MRR@10 than the most space-consuming one, while the former requires only 27 GiBs to represent the MS MARCO collection.

%% file: 5-conclusion.tex
\section{Conclusions}

In this paper, we introduced \sys{}, a novel ranking model that employs \textit{contextualized late interaction} over deep LMs (in particular, BERT) for efficient retrieval. By independently encoding queries and documents into fine-grained representations that interact via cheap and pruning-friendly computations, \sys{} can leverage the expressiveness of deep LMs while greatly speeding up query processing. In addition, doing so allows using \sys{} for end-to-end neural retrieval directly from a large document collection. Our results show that \sys{} is more than 170$\times$ faster and requires 14,000$\times$ fewer FLOPs/query than existing BERT-based models, all while only minimally impacting quality and while outperforming every non-BERT baseline.

\textbf{Acknowledgments.}
OK was supported by the Eltoukhy Family Graduate Fellowship at the Stanford School of Engineering. This research was supported in part by affiliate members and other supporters of the Stanford DAWN project---Ant Financial, Facebook, Google, Infosys, NEC, and VMware---as well as Cisco, SAP, and the NSF under CAREER grant CNS-1651570. Any opinions, findings, and conclusions or recommendations expressed in this material are those of the authors and do not necessarily reflect the views of the National Science Foundation.